\documentclass[a4paper,11pt]{article}
\usepackage{pos}
\usepackage{physics}

\title{General behavior of near-threshold hadron scattering for exotic hadrons}

\author*[a]{Katsuyoshi Sone}
\author[a]{Tetsuo Hyodo}

\affiliation[a]{Department of Physics, Tokyo Metropolitan University,\\ Hachioji 192-0397, Japan}

\emailAdd{sone-katsuyoshi@ed.tmu.ac.jp}
\emailAdd{hyodo@tmu.ac.jp}

\abstract{
We discuss the general behavior of the near-threshold scattering amplitude with channel couplings. The signal of the exotic hadrons near the threshold may manifest as a dip structure in the cross section originated from a zero point of the scattering amplitude. Such a dip structure by the zero point cannot be reproduced by the Flatt\'{e} amplitude which is widely used for the analysis of exotic hadrons, because of the constraints imposed on the Flatt\'{e} amplitude near the threshold. In this work, we propose the General amplitude which can describe the dip structure near the threshold, in contrast to the Flatt\'{e} amplitude. Moreover, we numerically study the behavior of the near-threshold cross section in relation to the zero point.

}

\FullConference{The XVIth Quark Confinement and the Hadron Spectrum Conference (QCHSC24)\\
 19-24 August, 2024\\
 Cairns Convention Centre, Cairns, Queensland, Australia\\}

\begin{document}
\maketitle

\section{Introduction}

Many exotic hadrons, such as $X(3872)$ and $f_0(980)$ emerge as sharp peak structures in the cross section near the threshold. The masses and decay widths of exotic hadrons are usually determined by fitting the peak structure using scattering amplitudes~\cite{Esposito:2021vhu, LHCb:2021auc}. 

On the other hand, several groups have reported that a dip structure appears instead of a peak structure in some systems where exotic hadrons are expected to exist~\cite{Dai:2014zta}. Such a dip structure can be caused by a zero point of the scattering amplitude~\cite{Dong:2020hxe, Baru:2024ptl}. 

Recently, the Flatt\'{e} amplitude~\cite{Flatte:1976xu} is often used for the analysis of near-threshold exotic states. However, the Flatt\'{e} amplitude has a limitation: the number of its parameters decreases near the threshold~\cite{Baru:2004xg}. Moreover, the Flatt\'{e} amplitude cannot reproduce the dip structures, because it does not have any zero point as discussed below~\cite{Sone:2024nfj}.

In this work, we propose the General amplitude which is more general than the Flatt\'{e} amplitude. We also investigate the behavior of the near-threshold cross sections numerically with the General amplitude focusing on its pole and zero point.

\section{Comparison of the Contact and Flatt\'{e} amplitudes} \label{sec: Conparison Eng}

In this section, we compare the Flatt\'{e} amplitude with the scattering amplitude derived from non-relativistic effective field theory with a contact interaction~\cite{Cohen:2004kf}. Through this comparison, we show that an additional condition is imposed on the Flatt\'{e} amplitude near the threshold on top of the unitarity.

We consider the two-body scattering with two channels where the threshold energy of channel 2 is higher than that of channel 1. In this study, we focus on the near-threshold energy region of channel 2. From the non-relativistic effective field theory with the contact interaction, the scattering amplitude (Contact amplitude) near the threshold of channel 2 is given by
\begin{align}
    f^{\rm C}(E) &= \frac{1}{\frac{1}{a_{12}^2} - \left(\frac{1}{a_{11}} + ip_0\right)\left(\frac{1}{a_{22}} + ik(E)\right)} 
    \begin{pmatrix}
        \frac{1}{a_{22}} + ik(E) & \frac{1}{a_{12}} \\
        \frac{1}{a_{12}} & \frac{1}{a_{11}} + ip_0
    \end{pmatrix},
    \label{eq:EFT amplitude Eng}
\end{align}
where $p_0$ and $k(E)$ represent the relative momenta of channels 1 and 2, respectively. $k(E)$ depends on the energy $E$, while $p_0$ is a constant, because we consider the near-threshold energy region of channel 2.  From Eq.~\eqref{eq:EFT amplitude Eng}, one see that the amplitude $f^{\rm C}(E)$ up to first order in $k$ is  characterized by three parameters, $a_{11},a_{12}$ and $a_{22}$.

On the other hand, the Flatt\'{e} amplitude is given as
\begin{align}
    f^{\rm F}(E) &= \frac{1}{2E_{\rm BW} - 2E - ig_1^2p_0 - ig_2^2k(E)} 
    \begin{pmatrix}
        g_1^2 & g_1g_2 \\
        g_1g_2 & g_2^2
    \end{pmatrix}, \label{eq: the Flatte derived from EFT Eng}
\end{align}
where $g_1$ and $g_2$ are the coupling constants of channels 1 and 2, and $E_{\rm BW}$ the bare energy. In order to compare the Flatt\'{e} and Contact amplitudes in the near-threshold energy region of channel 2, we neglect the $E\propto k^2$ term in Eq.~\eqref{eq: the Flatte derived from EFT Eng}:
\begin{align}
    f^{\rm F}(E) &\simeq \frac{1}{\frac{\alpha}{R} p_0 - i\frac{1}{R}p_0 - ik(E)}
    \begin{pmatrix}
        \frac{1}{R} & \sqrt{\frac{1}{R}} \\
        \sqrt{\frac{1}{R}} & 1
    \end{pmatrix},
    \label{eq:the Flatte amplitude up to first order in k which is represented by R and alpha Eng}
\end{align}
with $R=g_2^2/g_1^2,\ \alpha=2E_{\rm BW}/(g_1^2 p_0)$. From Eq.~\eqref{eq:the Flatte amplitude up to first order in k which is represented by R and alpha Eng}, we observe that the Flatt\'{e} amplitude near the threshold of channel 2 is determined by only two parameters~\cite{Baru:2004xg}.

Near the threshold of channel 2, while the Contact amplitude depends on three parameters, the Flatt\'{e} amplitude has only two parameters. It is known that the Contact amplitude has the same form as the amplitude given by K-matrix or M-matrix approach~\cite{Badalian:1981xj} which satisfies only the unitary condition. From these facts, we find that an additional condition is imposed on the Flatt\'{e} amplitude near the threshold.

Moreover, we compare the two amplitudes in terms of matrix inversion. The Flatt\'{e} amplitude does not take an invertible form, because the numerator is a rank-one matrix. On the other hand, the Contact amplitude is invertible as long as $a_{11},a_{12}$ and $a_{22}$ are finite. This difference indicate that, the Contact amplitude cannot be reduced to the Flatt\'{e} amplitude even if conditions are imposed on finite $a_{11},a_{12}$ and $a_{22}$.

\section{General amplitude} \label{sec: General amplitude Eng}

In this section, by modifying the parametrization of the Contact amplitude, we construct a scattering amplitude (General amplitude) that can be reduced to both the Flatt\'{e} and Contact amplitudes~\cite{Sone:2024nfj}. For this purpose, we introduce the real parameters  $A_{22}, \epsilon$ and $\gamma$ defined as
\begin{align}
    a_{11} = A_{22}\gamma, \quad
    a_{12} = \frac{A_{22}\gamma}{\sqrt{\epsilon-\gamma}}, \quad
    a_{22} = \frac{A_{22}\gamma}{\epsilon}. \label{eq: aij Eng}
\end{align}
Substituting these into Eq.~\eqref{eq:EFT amplitude Eng},  we obtain the General amplitude
\begin{align}
    f^{\rm G}(E) 
    &= \frac{1}{-\frac{1}{A_{22}} - i\epsilon p_0 - ik + A_{22} \gamma p_0k}
    \begin{pmatrix}
        \epsilon + iA_{22}\gamma k & \sqrt{\epsilon-\gamma} \\
        \sqrt{\epsilon-\gamma} &  1 + iA_{22}\gamma p_0
    \end{pmatrix}
    \label{eq:the general amplitude Eng}
\end{align}
Here, $\epsilon$ and $\gamma$ are dimensionless constants subject to the condition $\epsilon\geq \gamma$ to ensure the unitary. $A_{22}$ in units of length represents the scattering length of channel 2 in the absence of channel coupling ($\gamma=\epsilon$). The case with $A_{22}=0$ corresponds to the trivial scattering where all the components of $f^{\rm G}(E)$ vanish. Therefore, in the following, we consider only the cases with finite $A_{22}$.

Next, we show that the General amplitude can be reduced to both the Flatt\'{e} and Contact amplitudes. From Eq.~\eqref{eq: aij Eng}, we observe that $a_{11},a_{12}$ and $a_{22}$ remain finite as long as $\gamma\neq 0$. Therefore, when $\gamma$ is nonzero, $f^{\rm G}(E)$ is equivalent to the Contact amplitude~\eqref{eq:EFT amplitude Eng}. On the other hand, substituting $\gamma=0$ into Eq.~\eqref{eq:the general amplitude Eng}, we obtain
\begin{align}
    f^{\rm G}(E;A_{22}, \epsilon, \gamma=0) = \frac{1}{-\frac{1}{A_{22}} - i\epsilon p_0 - ik }
    \begin{pmatrix}
        \epsilon & \sqrt{\epsilon} \\
        \sqrt{\epsilon} &  1
    \end{pmatrix}. \label{eq: Flatte fromfG Eng}
\end{align}
Equation~\eqref{eq: Flatte fromfG Eng} shows that $f^{\rm G}(E)$ can be reduced to the Flatt\'{e} amplitude~\eqref{eq:the Flatte amplitude up to first order in k which is represented by R and alpha Eng} up to first order in $k$ with $\epsilon = 1/R,A_{22} = - R/(\alpha p_0)$.

Moreover, we examine the inverse matrix of the General amplitude in the case with $\gamma=0$. From Eq.~\eqref{eq:the general amplitude Eng}, the determinant of the General amplitude is given by
\begin{align}
    \det[f^{\rm G}(E)] 
    &= \frac{A_{22}\gamma}
    {\frac{1}{A_{22}}
    + ik
    + i\epsilon p_0
    - A_{22}\gamma p_0k}.
    \label{eq: det of fG}
\end{align}
Equation~\eqref{eq: det of fG} shows that $\det[f^{\rm G}(E)]$ becomes zero  when $\gamma=0$ and therefore $f^{\rm G}(E)$ does not have an inverse matrix. This is consistent with the fact that the Flatt\'{e} amplitude $(\gamma=0)$ is also not invertible.  From these results, we conclude that the General amplitude corresponds to the Contact amplitude when $\gamma\neq 0$, and it reduces to the Flatt\'{e} amplitude when $\gamma=0$.

\section{Pole and zeoro} \label{sec: cdd zero and pole Eng}

In this section, we discuss the pole and zero of the General amplitude. The pole of the scattering amplitude corresponds to the momentum at which the denominator of the amplitude vanishes. From Eq.~\eqref{eq:the general amplitude Eng}, the pole position of the General amplitude $k_{\rm p}^{\rm G}$ is determined by 
\begin{align}
    k_{\rm p}^{\rm G} &= i/a_{\rm G}, 
    \quad
    a_{\rm G} \equiv A_{22} \left(\frac{\frac{1}{A_{22}} + i\gamma p_0}{\frac{1}{A_{22}} + i\epsilon p_0}\right), 
    \label{eq:the scattering length of the new amplitude Eng}
\end{align}
where $a_{\rm G}$ represents the scattering length of channel 2 in the General amplitude.  Then, the pole of $f^{\rm G}(E)$ is determined only by the scattering length $a_{\rm G}$, because we approximate $f^{\rm G}(E)$ up to the first order in $k$. In general, $a_{\rm G}$ is complex because of the channel coupling effects. It is known that a pole at $k_{\rm p}^{\rm G}$ with $\Re[a_{\rm G}]>0$ $(\Re[a_{\rm G}]<0)$ corresponds to the quasibound (quasivirtual) state~\cite{Nishibuchi:2023acl}.

Next, we discuss the zero point of the amplitude which can lead to a dip structure near the threshold. In contrast to the pole position, each component of the amplitude generally has a different  zero point. It can be seen from Eq.~\eqref{eq:the general amplitude Eng} that the denominator of $f^{\rm G}(E)$ does not diverge, and therefore the zero points of the numerator determine the zero points of the amplitude~\cite{Baru:2024ptl}. Moreover, because only the $f_{11}^{\rm G}(E)$ component depends on the momentum $k$ in the numerator, only $f_{11}^{\rm G}(E)$ can have a zero point among four components in $f^{\rm G}(E)$. From Eq.~\eqref{eq:the general amplitude Eng}, we obtain the zero point of the General amplitude $k_{\rm zero}^{\rm G}$
\begin{align}
    k^{\rm G}_{\rm zero} &=  \frac{i}{A_{22}}\frac{\epsilon} {\gamma}.
    \label{eq:zero point of the fG11 component interms of k Eng}
\end{align}
Because the parameters $A_{22},\epsilon$ and $\gamma$ are given as real, the zero point $k_{\rm zero}^{\rm G}$ is pure imaginary. The corresponding zero point $E_{\rm zero}^{\rm G}$  in the complex energy plane emerges in the first (second) Riemann sheet when $\Im[k^{\rm G}_{\rm zero}]>0$ $(\Im[k^{\rm G}_{\rm zero}]<0)$. In addition, $E_{\rm zero}^{\rm G}$ lies on the negative real axis of $E$ (below the threshold of channel 2) because $k_{\rm zero}^{\rm G}$ is pure imaginary. A zero point with $\Im[k^{\rm G}_{\rm zero}]>0$ directly affects the physical scattering, because the scattering with $E<0$ occurs on the first sheet of channel 2. On the other hand, the Flatt\'{e} amplitude~\eqref{eq: Flatte fromfG Eng} does not have any zero points, because the numerator of $f^{\rm F}(E)$ is independent of $k$. This is consistent with the fact that $|k^{\rm G}_{\rm zero}|\to\infty$ with $\gamma\to0$ corresponding to the Flatt\'{e} amplitude.


\section{Numerical results} \label{sec: num results Eng}

In this section, we study the behavior of the cross section numerically using the General amplitude. Because the s-wave cross section is proportional to $|f^{\rm G}_{11}(E)|^2$, we use the normalized cross section~\cite{Sone:2024nfj} at the threshold of channel 2:
\begin{align}
    \sigma_{11}^{\rm N}(E) &\equiv |f^{\rm G}_{11}(E)|^2 / |f^{\rm G}_{11}(0)|^2
    = \left|\frac{1+i\frac{A_{22}\gamma}{\epsilon}k}{1+ia_{\rm G}k}\right|^2.
\end{align}
For the numerical calculation, we consider the $\pi\pi$-$K\bar{K}$ system with $f_0(980)$ as an example. We use the relativistic form of the momenta $p_0,k(E)$, because the threshold energy of $K\bar{K}$ is significantly higher that that of $\pi\pi$. The hadron masses are taken from PDG~\cite{ParticleDataGroup:2024cfk}.

In this analysis, we calculate $\sigma_{11}^{\rm N}(E)$ while varying $\gamma$ for a fixed scattering length $a_{\rm G}$. We note that the pole position $k_{\rm p}^{\rm G}$ remains unchanged under this condition, due to Eq.~\eqref{eq:the scattering length of the new amplitude Eng}. Since $\Re[a_{\rm G}]$ and $\Im[a_{\rm G}]$ are fixed, the two constraints are imposed on the independent three parameters of the General amplitude $A_{22},\epsilon$ and $\gamma$. Therefore, $f^{\rm G}(E)$ is uniquely determined by setting $\gamma$ under these conditions.

We consider the case with a fixed scattering length
\begin{align}
    a_{\rm G} = +1.0 - i0.8 \quad {\rm fm}. \notag
\end{align}
In this case, the pole on the complex energy plane is given by $E^{\rm G}_{\rm p}  =  -0.014-i0.048\ {\rm GeV}$ and $f^{\rm G}(E)$ has a quasibound pole below the $K\bar{K}$ threshold corresponding to $f_0(980)$. In this calculation, we examine the cases with $\gamma=0.07,\ 0.00,\ -0.01,\ -10.0$ as examples. The corresponding values of $A_{22}$ and $\epsilon$ are shown in Table~\ref{table1 Eng}.

\begin{table}
\centering
\begin{tabular}{ccccc}
\hline \hline 
$\gamma$ & $A_{22}\ (\rm fm)$& $\epsilon$ 
\\ \hline
        $+0.07$ & $+2.45$ & $+0.31$  \\
        $0.00$  & $+1.64$ & $+0.20$ \\
        $-0.01$ & $+1.59$ & $+0.19$  \\
        $-10.0$ & $+0.27$ & $-1.42$ \\
        \hline \hline 
\end{tabular}
\caption{
Parameters $A_{22}$ and $\epsilon$ for $a_{\rm G}=+1.0-i0.8$ fm.
}
\label{table1 Eng}
\end{table}

\begin{figure}[tbp]
    \centering
    \includegraphics[width = 8cm, clip]{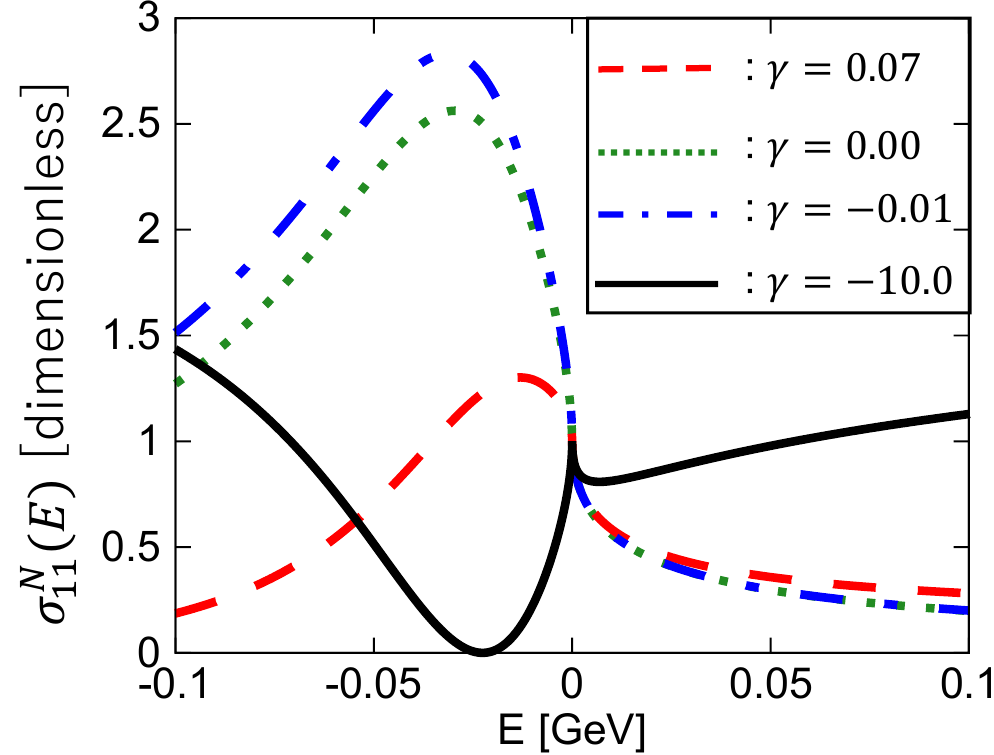}
    \caption{
    The cross sections $\sigma_{11}^{\rm N}(E)=|f^{\rm G}_{11}(E)|^2/|f^{\rm G}_{11}(0)|^2$ as functions of the energy $E$. The scattering length is fixed as $a_{\rm G}=+1.0-i1.0\ {\rm fm}$. The parameter is $\gamma=0.07$ (dashed line), $\gamma=0.00$ (dotted line), $\gamma=-0.01$ (dash-dotted line), and $\gamma=-10.0$ (solid line).
        }
    \label{fig:sigmaG11 A22-, epsilon-, aG=+1.0-i1.0 Eng}
\end{figure}

Normalized cross sections $\sigma_{11}^{\rm N}(E)$ for $\gamma=0.07,\ 0.00,\ -0.01,\ -10.0$ are shown in Fig. \ref{fig:sigmaG11 A22-, epsilon-, aG=+1.0-i1.0 Eng}. The dotted line ($\gamma=0$) corresponding to the Flatt\'{e} cross section exhibits a peak structure below the $K\bar{K}$ threshold. This peak is associated with the quasibound pole of $f_0(980)$. When $\gamma$ is slightly increased from zero, the height of the peak decreases and the position of the peak shifts toward the $K\bar{K}$ threshold, as shown by the dashed line ($\gamma=0.07$). On the other hand, when $\gamma$ is slightly decreased from zero, as shown by the dash-dotted line ($\gamma=-0.01$), the peak height becomes lager than that with $\gamma=0$. From these results, we find that $\sigma_{11}^{\rm N}(E)$ with $a_{\rm G}=+1.0-i1.0\ {\rm fm}$ exhibits a peak structure due to the quasibound pole for small $|\gamma|$. The dotted, dashed and dash-dotted lines can be fitted by the Flatt\'{e} amplitude, because they all show the peak structures. However, the scattering length obtained from the fitting of the cross sections with $\gamma\neq 0$ using the Flatt\'{e} amplitude may deviate from the correct value $a_{\rm G}=+1.0-i1.0\ {\rm fm}$, because those peak structures are affected by finite $\gamma$.

Next, we consider the case with $\gamma=-10.0$ (solid line). In this case, the cross section $\sigma_{11}^{\rm N}(E)$ exhibits a dip structure instead of a peak structure. This is caused  by the zero point discussed in Sec.~\ref{sec: cdd zero and pole Eng}. From Eq.~\eqref{eq:zero point of the fG11 component interms of k Eng} and Table~\ref{table1 Eng}, the position of the zero point on the complex energy plane is determined as
\begin{align}
    E^{\rm G}_{\rm zero} &=  
    -0.02\ {\rm GeV}, 
    \label{eq: Ep1}
\end{align}
where $E^{\rm G}_{\rm zero}$ is located on the real axis. In fact, $\sigma_{11}^{\rm N}(E)$ vanishes at $E^{\rm G}_{\rm zero}$ in Eq.~\eqref{eq: Ep1} causing the dip structure. Reference~\cite{Dai:2014zta} reported a dip structure below the $K\bar{K}$ threshold in the $\gamma \gamma \to \pi\pi$ reaction. This dip may originate from the zero point of the $\pi\pi$ scattering amplitude. When the cross section has a zero point near the threshold, as seen in the solid line ($\gamma=-10.0$), the Flatt\'{e} amplitude fails to describe this behavior, because the Flatt\'{e} amplitude does not have any zero points as discussed in Sec.~\ref{sec: cdd zero and pole Eng}.

For $\Im[k^{\rm G}_{\rm zero}]>0$, the zero point emerges in the physical scattering region, as discussed in Sec.~\ref{sec: cdd zero and pole Eng}. Table~\ref{table1 Eng} shows that the parameter sets satisfying $\Im[k^{\rm G}_{\rm zero}]>0$ correspond to the dashed ($\gamma=0.07$) and solid ($\gamma=-10.0$) lines in Fig.~\ref{fig:sigmaG11 A22-, epsilon-, aG=+1.0-i1.0 Eng}. Therefore, in addition to the solid line, the dashed line also processes a zero point in the physical scattering region. Here, Eq.~\eqref{eq:zero point of the fG11 component interms of k Eng} shows that $|k_{\rm zero}^{\rm G}|$ increases as $\gamma$ is decreased. In fact, for $\gamma=-10.0$ the zero point is located at $E^{\rm G}_{\rm zero}=-0.25\ \rm GeV$ which lies outside the range of  Fig.~\ref{fig:sigmaG11 A22-, epsilon-, aG=+1.0-i1.0 Eng}.

\section{Summary}

In this contribution, we discuss the general behavior of the scattering amplitude near the threshold using the General amplitude. First, in Sec.~\ref{sec: Conparison Eng}, we show that a constraint is imposed on the Flatt\'{e} amplitude near the threshold by comparing the Flatt\'{e} and Contact amplitudes. We also show that the Contact amplitude cannot be directly reduced to the Flatt\'{e} amplitude by simply imposing the condition on the parameters. 

Based on this, we introduce the new parametrization to construct the General amplitude which can be reduced to both the Flatt\'{e} and Contact amplitudes by varying the parameter $\gamma$ in Sec.~\ref{sec: General amplitude Eng}. In Sec.~\ref{sec: cdd zero and pole Eng}, we determine the positions of the pole and zero point of the General amplitude. Moreover, we show that  the Flatt\'{e} amplitude corresponding to the $\gamma=0$ case does not exhibit any zero point, because the zero point of the General amplitude goes to infinity with $\gamma\to0$. 

Finally, we numerically study the behavior of the near-threshold cross section by varying $\gamma$ while keeping the scattering length fixed, which results in the formation of a quasibound pole. As a result, the cross section derived from the General amplitude with the quasibound pole exhibits a peak structure below the threshold for small $|\gamma|$ similarly to the Flatt\'{e} amplitude. Conversely, for large negative $\gamma$, the cross section displays a dip structure below the threshold caused by a zero point rather than a peak structure. In this way, we show the shape of the cross section may significantly differ from the typical peak structures even if the scattering amplitude has a resonance pole near the threshold.


\begin{acknowledgments}
This work has been supported in part by the Grants-in-Aid for Scientific Research from JSPS (Grants
No.~JP23H05439, 
No. JP22K03637, and 
No. JP18H05402),
 by the RCNP Collaboration Research network (COREnet) 048 "Revealing the nature of exotic hadrons in Belle (II) by collaboration of experimentalists and theorists", 
and by JST SPRING, 
Grant Number JPMJSP2156

\end{acknowledgments}


\begin{thebibliography}{10}

\bibitem{Esposito:2021vhu}
A.~Esposito, L.~Maiani, A.~Pilloni, A.D.~Polosa and V.~Riquer, \emph{{From the
  line shape of the $X(3872)$ to its structure}},
  \href{https://doi.org/10.1103/PhysRevD.105.L031503}{\emph{Phys. Rev. D}
  {\bfseries 105} (2022) L031503}
  [\href{https://arxiv.org/abs/2108.11413}{{\ttfamily 2108.11413}}].

\bibitem{LHCb:2021auc}
{\scshape LHCb} collaboration, \emph{{Study of the doubly charmed tetraquark
  $T_{cc}^{+}$}},
  \href{https://doi.org/10.1038/s41467-022-30206-w}{\emph{Nature Commun.}
  {\bfseries 13} (2022) 3351}
  [\href{https://arxiv.org/abs/2109.01056}{{\ttfamily 2109.01056}}].

\bibitem{Dai:2014zta}
L.-Y.~Dai and M.R.~Pennington, \emph{{Comprehensive amplitude analysis of
  $\gamma\gamma \rightarrow \pi^+\pi^-, \pi^0\pi^0$ and $\overline{K} K$ below
  1.5 GeV}}, \href{https://doi.org/10.1103/PhysRevD.90.036004}{\emph{Phys. Rev.
  D} {\bfseries 90} (2014) 036004}
  [\href{https://arxiv.org/abs/1404.7524}{{\ttfamily 1404.7524}}].

\bibitem{Dong:2020hxe}
X.-K.~Dong, F.-K.~Guo and B.-S.~Zou, \emph{{Explaining the Many Threshold
  Structures in the Heavy-Quark Hadron Spectrum}},
  \href{https://doi.org/10.1103/PhysRevLett.126.152001}{\emph{Phys. Rev. Lett.}
  {\bfseries 126} (2021) 152001}
  [\href{https://arxiv.org/abs/2011.14517}{{\ttfamily 2011.14517}}].

\bibitem{Baru:2024ptl}
V.~Baru, F.-K.~Guo, C.~Hanhart and A.~Nefediev, \emph{{How does the $X(3872)$
  show up in $e^+e^-$ collisions: Dip versus peak}},
  \href{https://doi.org/10.1103/PhysRevD.109.L111501}{\emph{Phys. Rev. D}
  {\bfseries 109} (2024) L111501}
  [\href{https://arxiv.org/abs/2404.12003}{{\ttfamily 2404.12003}}].

\bibitem{Flatte:1976xu}
S.M.~Flatte, \emph{{Coupled - Channel Analysis of the $\pi \eta$ and $K\bar{K}$
  Systems Near $K\bar{K}$ Threshold}},
  \href{https://doi.org/10.1016/0370-2693(76)90654-7}{\emph{Phys. Lett. B}
  {\bfseries 63} (1976) 224}.

\bibitem{Baru:2004xg}
V.~Baru, J.~Haidenbauer, C.~Hanhart, A.E.~Kudryavtsev and U.-G.~Meissner,
  \emph{{Flatte-like distributions and the $a_0(980)/f_0(980)$ mesons}},
  \href{https://doi.org/10.1140/epja/i2004-10105-x}{\emph{Eur. Phys. J. A}
  {\bfseries 23} (2005) 523}
  [\href{https://arxiv.org/abs/nucl-th/0410099}{{\ttfamily nucl-th/0410099}}].

\bibitem{Sone:2024nfj}
K.~Sone and T.~Hyodo, \emph{{General amplitude of near-threshold hadron
  scattering for exotic hadrons}},
  \href{https://arxiv.org/abs/2405.08436}{{\ttfamily 2405.08436}}.

\bibitem{Cohen:2004kf}
T.D.~Cohen, B.A.~Gelman and U.~van Kolck, \emph{{An Effective field theory for
  coupled channel scattering}},
  \href{https://doi.org/10.1016/j.physletb.2004.03.020}{\emph{Phys. Lett. B}
  {\bfseries 588} (2004) 57}
  [\href{https://arxiv.org/abs/nucl-th/0402054}{{\ttfamily nucl-th/0402054}}].

\bibitem{Badalian:1981xj}
A.M.~Badalian, L.P.~Kok, M.I.~Polikarpov and Y.A.~Simonov, \emph{{Resonances in
  Coupled Channels in Nuclear and Particle Physics}},
  \href{https://doi.org/10.1016/0370-1573(82)90014-X}{\emph{Phys. Rept.}
  {\bfseries 82} (1982) 31}.

\bibitem{Nishibuchi:2023acl}
T.~Nishibuchi and T.~Hyodo, \emph{{Analysis of the $\Xi(1620)$ resonance and
  $\bar{K}\Lambda$ scattering length with a chiral unitary approach}},
  \href{https://doi.org/10.1103/PhysRevC.109.015203}{\emph{Phys. Rev. C}
  {\bfseries 109} (2024) 015203}
  [\href{https://arxiv.org/abs/2305.10753}{{\ttfamily 2305.10753}}].

\bibitem{ParticleDataGroup:2024cfk}
{\scshape Particle Data Group} collaboration, \emph{{Review of particle
  physics}}, \href{https://doi.org/10.1103/PhysRevD.110.030001}{\emph{Phys.
  Rev. D} {\bfseries 110} (2024) 030001}.

\end{thebibliography}

\providecommand{\href}[2]{#2}\begingroup\raggedright\endgroup

\end{document}